# Analytical solutions of the Schrödinger equation with Kratzer- screened Coulomb potential to the Quarkonium systems

Etido P. Inyang, Ephraim P. Inyang, Joseph E. Ntibi and Eddy S. William

*Theoretical Physics Group, Department of Physics, University of Calabar, PMB 1115, Calabar Nigeria*

Corresponding author email: etidophysics@gmail.com

**ABSTRACT**

In this work, we obtain the Schrödinger equation solutions for the Kratzer potential plus screened Coulomb potential model using the series expansion method. The energy eigenvalues is obtained in non-relativistic regime and the corresponding unnormalized eigenfunction. Three special cases were obtained. We applied the present results to calculate heavy-meson masses of charmonium $c\bar{c}$ and bottomonium $b\bar{b}$, and we got the numerical values for 1S, 2S, 1P,2P, 3S, 4S ,1D,2D and 1F states. The results are in good agreement with experimental data and the work of other researchers.

## 1. INTRODUCTION

The study of heavy quarkonium systems such as charmonium and bottomonium plays an important role in understanding the quantitative tests of quantum chromodynamics (QCD) and the standard model [1]. These systems can be studied within the Schrödinger equation (SE) [2]. The solution of SE with spherically symmetric potential is one of the important problems in physics and chemistry. It plays an important role for spectroscopy, molecules and nuclei, in particular, the properties of constituent's particles and dynamics of their interactions [3]. There potential should take into account the two important features of the strong interaction, namely,asymptotic freedom and quark confinement[4]. The SE has been solved using various methods such as, asymptotic iteration method (AIM)[5] Laplace transformation method [6], super symmetric quantum mechanics method (SUSQM)[7] , Nikiforov-Uvarov(NU) method [8-22] ,series expansion method [23] and others.

The most fundamental potential used in studying quarkonium system is the Cornell potential, also known as funnel potential. Most researchers have carried out works with Cornell potential. For instance, Vega and Flores [24] solved the Schrödinger equation with the Cornell potential using the variational method and super symmetric quantum mechanics (SUSYQM). Ciftci and Kisoglu [25] addressed non-relativistic arbitrary $l$ -states of quark-antiquark through the Asymptotic Iteration Method (AIM). The energy eigenvalues with any $l \neq 0$ states and mass of the massive quark-antiquark system (quarkonium) were gotten. An analytic solution of the N-dimensional radial Schrödinger equation with the mixture of vector and scalar potentials via the Laplace transformation method (LTM) was studied by [26]. Their results were employed to analyze the different properties of the heavy-light mesons. Al-Jamel and Widyan [27] studied heavy quarkonium ($c\bar{c}$ and $b\bar{b}$) mass spectra in a Coulomb field plus quadratic potential using the Nikiforov-Uvarov method. In their work, the spin-averaged mass spectra of heavy quarkonia ($c\bar{c}$ and $b\bar{b}$) in a Coulomb plus quadratic potential is analyzed within the non-relativistic Schrödinger equation. Al-Oun  et al**.** [28] examine heavy quarkonia ($c\bar{c}$ , and $b\bar{b}$) characteristics in the general framework of a non-relativistic potential model consisting of a Coulomb plus quadratic potential. Furthermore, Omugbe et al.[29] solved the SE with Killingbeck potential plus an inversely quadratic potential model. They obtained the energy eigenvalues and the mass spectra of the heavy and heavy-light meson systems. In addition, Al-Jamel, [30] studied the energy spectra of mesons and hadronic interactions using Numerov's method. Their solutions were used to describe the phenomenological interactions between the charm-anticharm quarks via the model. The model accurately predicts the mass spectra of charmed quarkonium as an example of mesonic systems. Inyang et al.[31] obtained the Klein-Gordon equation solutions for the Yukawa potential using the Nikiforov-Uvarov method. The energy eigenvalues were obtained both in relativistic and non-relativistic regime. They applied the results to calculate heavy-meson masses of charmonium $c\bar{c}$ and bottomonium $b\bar{b}$.

The Kratzer potential is one of the widely used potential models in molecular physics and quantum chemistry [32]. The Kratzer potential contains a repulsive part and long-range attraction. Apart from that, the potential is also known to approach infinity when the inter-nuclear distance approaches zero, due to the repulsion that exists between the molecules of the potential. As the inter-nuclear molecular distance approaches infinity, the potential decomposes to zero [33,34]. The potential is of the form

$$V(r) = a - \frac{b}{r} + \frac{c}{r^2} \qquad (1)$$

where $a$, $b$ and $c$ are potential strength parameters. The Kratzer potential has been used as a potential model to describe inter-nuclear vibration of diatomic molecules by different authors [35, 36]. Another potential model used significantly in nuclear, particle and condensed matter physics is the screened Coulomb potential, also known as Yukawa potential [37]. The screened Coulomb potential is a short-range potential [38]. In solid-state physics, it describes the charge particle effects of conduction electrons [39]. It takes the form

$$V(r) = -\frac{pe^{-\alpha r}}{r}, \qquad (2)$$

where $p$ is the screened Coulomb potential parameter, $\alpha$ is the screening parameter and $r$ is the distance between two particles.

Many researchers, in recent time are concerned with combining two or more potentials. The fundamental nature of combining two or more physical potential models is to have a broader range of applications [40]. For example, Cornell potential, which is the combination of Coulomb potential with linear terms, is used in studying the mass spectra for coupled states and the electromagnetic characteristics of meson [41]. With this in mind, we attempt to solve the SE with a potential obtained from the combination of Kratzer potential [Eq.(1)] and screened Coulomb potential [ Eq.(2)] using the series expansion method to obtain the mass spectra of heavy quarkonium systems. To the best of our knowledge this is the first time Kratzer - screened Coulomb potential model is being studied with the aim of determining the mass spectra of heavy quarkoniun systems.

The combine potential takes the form

$$V(r) = a - \frac{b}{r} + \frac{c}{r^2} - \frac{pe^{-\alpha r}}{r}. \tag{3}$$

When we set $p = 0$, Eq.(3) reduces to Kratzer potential. Also, when $a = c = p = \alpha = 0$, Eq.(3) reduces to standard Coulomb potential, when $a = b = c = 0$, Eq.(3) reduces to screened Coulomb potential. We organized this paper as follows: in section 2, we shall focus primarily on SE solution for the Kratzer- screened Coulomb potential system using the series expansion method. In section 3, we shall discuss special cases of potential. In section 4, we discuss the results, and in section 5, we give a concluding remark.

## 2. BOUND STATE SOLUTIONS TO THE SCHRÖDINGER EQUATION WITH KRATZER-SCREENED COULOMB POTENTIAL

We consider the radial SE of the form [23]

$$\frac{d^2 R(r)}{dr^2} + \frac{2}{r}\frac{dR(r)}{dr} + \left[\frac{2\mu}{\hbar^2}(E - V(r)) - \frac{l(l+1)}{r^2}\right] R(r) = 0, \tag{4}$$

where $l$ is rotational quantum number taking the values $0,1,2,3,4…,$ $\mu$ is the reduced mass, $r$ is the internuclear separation and, $E$ denotes the energy eigenvalues of the system.

We carry out series expansion of the exponential term in Eq. (3) up to order three, in order to make the potential to interact in the quark-antiquark system and this yields,

$$\frac{e^{-\alpha r}}{r} = \frac{1}{r} - \alpha + \frac{\alpha^2 r}{2} - \frac{\alpha^3 r^2}{6} + … \tag{5}$$

By substituting Eq.(5) into Eq.(3) we obtain

$$V(r) = \frac{\alpha_0}{r^2} - \frac{\alpha_1}{r} + \alpha_2 r + \alpha_3 r^2 + \alpha_4, \tag{6}$$

where

$$\alpha_0 = c, \quad \alpha_1 = b + p, \quad \alpha_2 = \frac{p\alpha^2}{2}, \quad \alpha_3 = -\frac{p\alpha^3}{6}, \quad \alpha_4 = a + p\alpha \bigg\}. \tag{7}$$

We substitute Eq.(6) into Eq.(4) and obtain

$$\frac{d^2R(r)}{dr^2} + \frac{2}{r}\frac{dR(r)}{dr} + \left[\varepsilon + \frac{A}{r} - Br - Cr^2 - \frac{L(L+1)}{r^2}\right]R(r) = 0, \tag{8}$$

where

$$\varepsilon = \frac{2\mu}{\hbar^2}(E - \alpha_4), \ A = -\frac{2\mu\alpha_1}{\hbar^2}, \ B = \frac{2\mu\alpha_2}{\hbar^2}, \ C = \frac{2\mu\alpha_3}{\hbar^2}, \tag{9}$$

$$L(L+1) = \frac{2\mu\alpha_0}{\hbar^2} + l(l+1). \tag{10}$$

From Eq.(10) we have that

$$L = -\frac{1}{2} + \frac{1}{2}\sqrt{(2l+1)^2 + \frac{8\mu\alpha_0}{\hbar^2}}. \tag{11}$$

Now make an anzats wave function [42]

$$R(r) = e^{-\alpha r^2 - \beta r}F(r), \tag{12}$$

where $\alpha$ and $\beta$ are positive constants. Differentiating Eq.(12) twice we obtain the following:

$$R'(r) = F'(r)e^{-\alpha r^2 - \beta r} + F(r)(-2\alpha r - \beta)e^{-\alpha r^2 - \beta r}, \tag{13}$$

$$R''(r) = F''(r)e^{-\alpha r^2 - \beta r} + F'(r)(-2\alpha r - \beta)e^{-\alpha r^2 - \beta r} + \left[(-2\alpha) + (-2\alpha r - \beta)(-2\alpha r - \beta)\right]F(r)e^{-\alpha r^2 - \beta r}. \tag{14}$$

Substituting Eqs. (12), (13) and (14) into Eq.(8) and divide through by $e^{-\alpha r^2 - \beta r}$ we obtain

$$F''(r) + \left[-4\alpha r - 2\beta + \frac{2}{r}\right]F'(r) + \left[\begin{array}{c}(4\alpha^2 - C)r^2 + (4\alpha\beta - B)r \\ +(A - 2\beta)\frac{1}{r} - \frac{L(L+1)}{r^2} + (\varepsilon + \beta^2 - 6\alpha)\end{array}\right]F(r) = 0. \tag{15}$$

The function $F(r)$ is a series of the form,

$$F(r) = \sum_{n=0}^{\infty} a_n r^{2n+L}. \tag{16}$$

Taking the first and second derivatives of Eq. (16), we obtain the following:

$$F'(r) = \sum_{n=0}^{\infty}(2n+L)a_n r^{2n+L-1}, \tag{17}$$

$$F''(r) = \sum_{n=0}^{\infty}(2n+L)(2n+L-1)a_n r^{2n+L-2}. \tag{18}$$

We substitute for Eqs. (16), (17) and (18) into Eq.(15) and obtain

$$\sum_{n=0}^{\infty}(2n+L)(2n+L-1)a_n r^{2n+L-2} + \left[-4\alpha r - 2\beta + \frac{2}{r}\right]\sum_{n=0}^{\infty}(2n+L)a_n r^{2n+L-1}$$
$$+ \left[(4\alpha^2 - C)r^2 + (4\alpha\beta - B)r + \frac{(A - 2\beta)}{r} - \frac{L(L+1)}{r^2} + (\varepsilon + \beta^2 - 6\alpha)\right]\sum_{n=0}^{\infty}a_n r^{2n+L} = 0. \tag{19}$$

By collecting powers of $r$ in Eq.(19) we have

$$\sum_{n=0}^{\infty} a_n \left\{ \begin{array}{l} \left[(2n+L)(2n+L-1)+2(2n+L)-L(L+1)\right]r^{2n+L-2} + \left[-2\beta(2n+L)+(A-2\beta)\right]r^{2n+L-1} \\ +\left[-4\alpha(2n+L)+\varepsilon+\beta^2-6\alpha\right]r^{2n+L} + \left[4\alpha\beta-B\right]r^{2n+L+1} + \left[4\alpha^2-C\right]r^{2n+L+2} \end{array} \right\} = 0 \quad (20)$$

Equation (20) is linearly independent, implying that each of the terms is separately equal to zero, noting that $r$ is a non-zero function; therefore, it is the coefficient of $r$ that is zero. With this in mind, we obtain the relation for each of the terms.

$$(2n+L)(2n+L-1)+2(2n+L)-L(L+1) = 0 \quad (21)$$

$$-2\beta(2n+L)+A-2\beta = 0 \quad (22)$$

$$-4\alpha(2n+L)+\varepsilon+\beta^2-6\alpha = 0 \quad (23)$$

$$4\alpha\beta - B = 0 \quad (24)$$

$$4\alpha^2 - C = 0 \quad (25)$$

From Eq.(22) we have,

$$\beta = \frac{A}{4n+2L+2}. \quad (26)$$

From Eq.(25) we have,

$$\alpha = \frac{\sqrt{C}}{2}. \quad (27)$$

We proceed to obtain the energy eigenvalue equation using Eq.(23) and have

$$\varepsilon = 2\alpha(4n+2L+3) - \beta^2 \quad (28)$$

Substituting Eqs. (9),(11),(26) and (27) into Eq.(28) and simplifying we obtain

$$E_{nl} = \sqrt{\frac{\hbar^2 \alpha_3}{2\mu}} \left(4n+2+\sqrt{(2l+1)^2+\frac{8\mu\alpha_0}{\hbar^2}}\right) - \frac{2\mu\alpha_1^2}{\hbar^2}\left(4n+1+\sqrt{(2l+1)^2+\frac{8\mu\alpha_0}{\hbar^2}}\right)^{-2} + \alpha_4 \quad (29)$$

Substituting Eq.(7) into Eq.(29) we obtain the energy eigenvalue for the combined potential of Eq.(3) as;

$$E_{nl} = \sqrt{\frac{-\hbar^2 p\alpha^3}{12\mu}} \left(4n+2+\sqrt{(2l+1)^2+\frac{8\mu c}{\hbar^2}}\right) - \frac{2\mu}{\hbar^2}(-b-p)^2\left(4n+1+\sqrt{(2l+1)^2+\frac{8\mu c}{\hbar^2}}\right)^{-2} + a + p\alpha \quad (30)$$

Upon substituting Eqs (9),(11),(16),(26) and (27) into Eq.(12) we obtain the unnormalized wave function in the form

$$R(r) = \sum_{n=0}^{\infty} a_n r^{2n-\frac{1}{2}+\frac{1}{2}\sqrt{(2l+1)^2+\frac{8\mu c}{\hbar^2}}} e^{-\frac{\sqrt{\frac{-\alpha^3\mu p}{3\hbar^2}}}{2}r^2 - \left[\frac{2\mu(-b-p)}{\hbar^2\left(4n-1+\sqrt{(2l+1)^2+\frac{8\mu c}{\hbar^2}}\right)}\right]r} \quad (31)$$

where

$a_n$ = normalization constant

## 3. SPECIAL CASES

1. Setting $p = 0$ in Eq.(30) we obtain the energy eigenvalue for Kratzer potential in the form

$$E_{nl} = \left(4n + 2 + \sqrt{(2l+1)^2 + \frac{8\mu c}{\hbar^2}}\right) - \frac{2\mu b^2}{\hbar^2}\left(4n + 1 + \sqrt{(2l+1)^2 + \frac{8\mu c}{\hbar^2}}\right)^{-2} + a \qquad (32)$$

2. Setting $p = a = c = \alpha = 0$ in Eq.(30) we obtain the energy eigenvalue for standard Coulomb potential in the form

$$E_{nl} = -\frac{2\mu b^2}{\hbar^2}\left(4n + 1 + \sqrt{(2l+1)^2}\right)^{-2} \qquad (33)$$

3. Setting $a = b = c = 0$ in Eq.(30) we obtain the energy eigenvalue for screened Coulomb potential in the form

$$E_{nl} = \sqrt{\frac{-\hbar^2 p\alpha^3}{12\mu}}\left(4n + 2 + \sqrt{(2l+1)^2}\right) - \frac{2\mu p^2}{\hbar^2}\left(4n + 1 + \sqrt{(2l+1)^2}\right)^{-2} + p\alpha \qquad (34)$$

## 4. RESULTS AND DISCUSSION

4.1 RESULTS

The mass spectra of the heavy quarkonium system such as charmonium and bottomonium that have the quark and antiquark flavor is calculated and we apply the following relation [43, 44]

$$M = 2m + E_{nl},$$
(35)

where $m$ is quarkonium bare mass, and $E_{nl}$ is energy eigenvalues. By substituting Eq. (30) into Eq. (35) we obtain the mass spectra for Kratzer-screen Coulomb potential as:

$$M = 2m + \sqrt{\frac{-\hbar^2 p\alpha^3}{12\mu}}\left(4n + 2 + \sqrt{(2l+1)^2 + \frac{8\mu c}{\hbar^2}}\right) - \frac{2\mu}{\hbar^2}(-b-p)^2\left(4n + 1 + \sqrt{(2l+1)^2 + \frac{8\mu c}{\hbar^2}}\right)^{-2} + a + p\alpha \qquad (36)$$

**Table 1.** Mass spectra of charmonium in (GeV) ($m_c = 1.488$ GeV, $\mu = 0.744$ GeV, $\alpha = 3.1674$, $\hbar = 1$, $a = -0.2860$ GeV, $b = 0.001$ GeV, $c = 0.1306$ GeV and $P = 0.0022$ GeV)

| State | Present work | [44] | [31] | Experiment[46] |
|---|---|---|---|---|
| 1S | 3.096 | 3.096 | 3.096 | 3.096 |
| 2S | 3.686 | 3.686 | 3.686 | 3.686 |
| 1P | 3.295 | 3.255 | 3.527 | 3.525 |
| 2P | 3.802 | 3.779 | 3.687 | 3.773 |
| 3S | 4.040 | 4.040 | 4.040 | 4.040 |
| 4S | 4.269 | 4.269 | 4.360 | 4.263 |
| 1D | 3.583 | 3.504 | 3.098 | 3.770 |
| 2D | 3.976 | - | 3.976 | 4.159 |
| 1F | 3.862 | - | 4.162 | - |

**Table 2.** Mass spectra of bottomonium in (GeV) ($m_b = 4.680$ GeV, $\mu = 2.340$ GeV, $\alpha = 4.4477$, $\hbar = 1$, $a = -0.0273$ GeV, $b = 0.001$ GeV, $c = 0.050$ GeV and $P = 0.0022$ GeV)

| State | Present work | [44] | [31] | Experiment[46] |
|---|---|---|---|---|
| 1S | 9.460 | 9.460 | 9.460 | 9.460 |
| 2S | 10.569 | 10.023 | 10.023 | 10.023 |
| 1P | 9.661 | 9.619 | 9.661 | 9.899 |
| 2P | 10.138 | 10.114 | 10.238 | 10.260 |
| 3S | 10.355 | 10.355 | 10.355 | 10.355 |
| 4S | 10.567 | 10.567 | 10.567 | 10.580 |
| 1D | 9.943 | 9.864 | 9.943 | 10.164 |
| 2D | 10.306 | - | 10.306 | - |
| 1F | 10.209 | - | 10.209 | - |

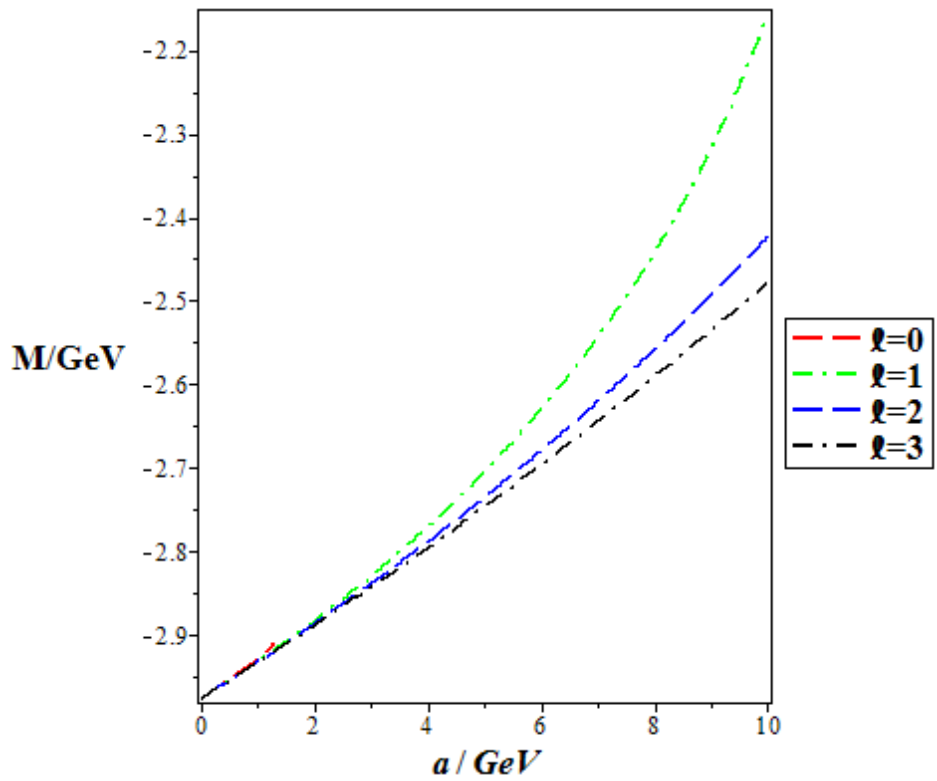

Figure 1. Variation of mass spectra with potential strength $(a)$ for different quantum numbers

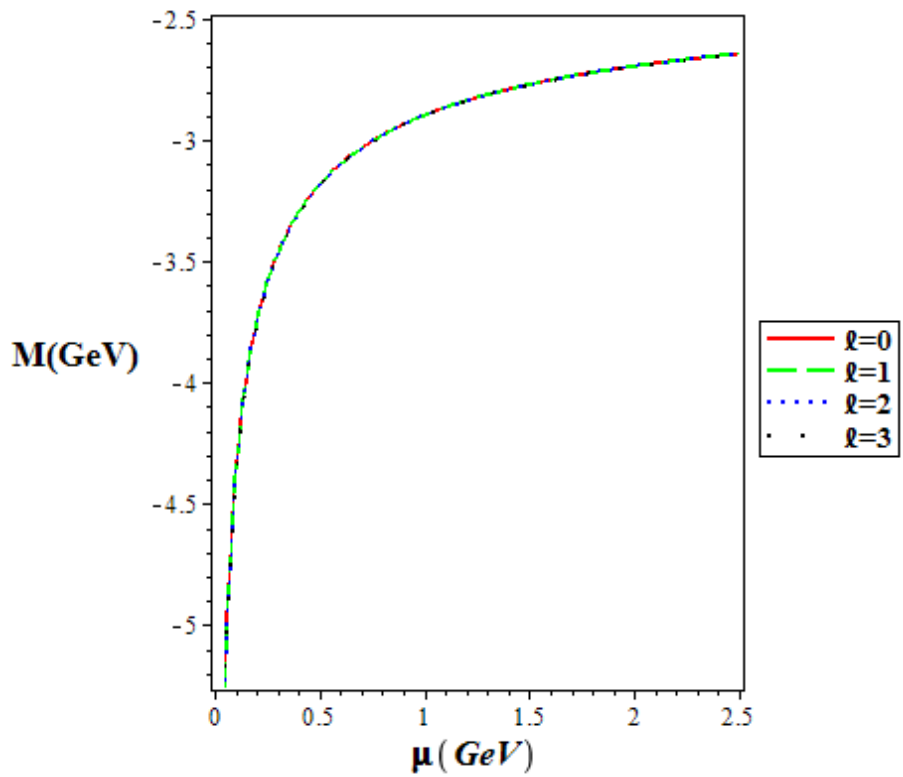

Figure 2. Variation of mass spectra with reduced mass $\mu$ for different quantum numbers

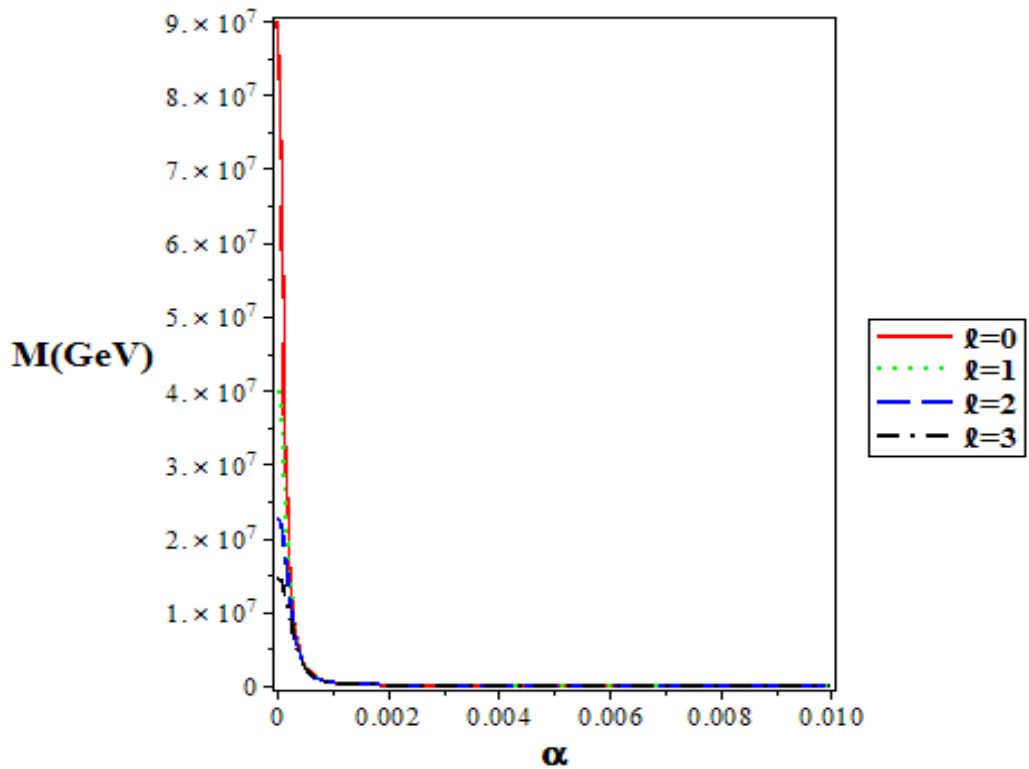

Figure 3. Variation of mass spectra with screening parameter ($\alpha$) for different quantum number

4.2 Discussion of results

We calculate mass spectra of charmonium and bottomonium for states from 1S, 2S, 1P,2P, 3S, 4S ,1D,2D and 1F by using Eq. (36). We adopt the numerical values of bottomonium $(b\bar{b})$ and charmonium $(c\bar{c})$ masses as 4.68 $GeV$ and 1.488 $GeV$, respectively, Ref. [45]. Then, the corresponding reduced mass are $\mu_b$ = 2.340 $GeV$ and $\mu_c = 0.744\, GeV$. The free parameters of Eq.(36) were then gotten by solving two algebraic equations by inserting experimental data of mass spectra for $2S, 2P$ in the case of charmonium. In the case of bottomonium the values of the free parameters in Eq. (36) are calculated by solving two algebraic equations, which were obtained by inserting experimental data of mass spectra for $1S, 2S$. Experimental data is taken from Ref. [46].

We note that calculation of mass spectra of charmonium and bottomonium are in good agreement with experimental data and other theoretical calculations. The values obtained are in a good agreement with the work of other researchers like in Ref.[44] and Ref.[31], as shown in tables 1 and 2. In Ref.[44] the author investigated the N-radial SE analytically by employing Cornell potential, which was extended to finite temperature. In Ref.[31] the Klein-Gordon equation is solved for the Yukawa potential using the Nikiforov-Uvarov method. The energy eigenvalues were obtained both in relativistic and non-relativistic regime. The results were used to calculate heavy-meson masses of charmonium $c\bar{c}$ and bottomonium $b\bar{b}$. We also plotted mass spectra energy against potential strength $(a)$, reduced mass $(\mu)$ and screening parameter $(\alpha)$ respectively. In Fig. 1, the mass spectra energy converges at the beginning but spread out, and there is a monotonic increase in potential strength $(a)$. Figures 2 and 3 shows the convergence of the mass spectra energy as the screening parameter $(\alpha)$ and reduced mass $(\mu)$ increases for various angular quantum numbers.

5. Conclusion

In this work, we have obtained the bound state solutions of the Schrödinger equation for the Kratzer plus screened Coulomb potential via the series expansion method. The energy eigenvalues are obtained in a non-relativistic regime. The corresponding unnormalized eigenfunction was also obtained. We applied the present results to calculate heavy-meson masses such as charmonium and bottomonium. The mass spectra energy of charmonium $(c\bar{c})$, and bottomonium $(b\bar{b})$ for states 1S to 1F were obtained and compared with experimental data and other theoretical works, which are in good agreement. We plotted the mass spectra energy against potential strength, screening parameter and reduced mass respectively. The energy eigenvalues can be used to study the suitability of material for mixed radiation dosimety as in Ref. [47]. The analytical solutions can also be used to describe other characteristics of the quarkonium systems like thermodynamic properties.

**DECLARATIONS:**

**AVAILABILITY OF DATA AND MATERIALS**
All data generated during this study are included in the references in the paper.

**COMPETING INTERESTS**
The authors declare that they have no competing interests.

**FUNDING**
Not applicable

**AUTHORS CONTRIBUTIONS**
Dr. E. P. Inyang suggested the point research and carried out the calculations and wrote it. Dr.E.P.Inyang carried out the results and reviewed it. Dr. J.E. Ntibi follows up with writing the literature. Dr. E.S.William carried out the writing of the full manuscript. All authors read and approved the final manuscript.

**ACKNOWLEGEMENTS**
Dr. E.P.Inyang wish to thank Prof. A. N. Ikot, Department of Physics, University of South Africa for his encouragement for the successful completion of this work.